\documentclass[%
 reprint,
 amsmath, assume,
 aps,
]{revtex4-2}

\usepackage{pdfpages} 
\usepackage{pgffor} 

\makeatletter
\AtBeginDocument{\let\LS@rot\@undefined}
\makeatother

\def\supplementfilename{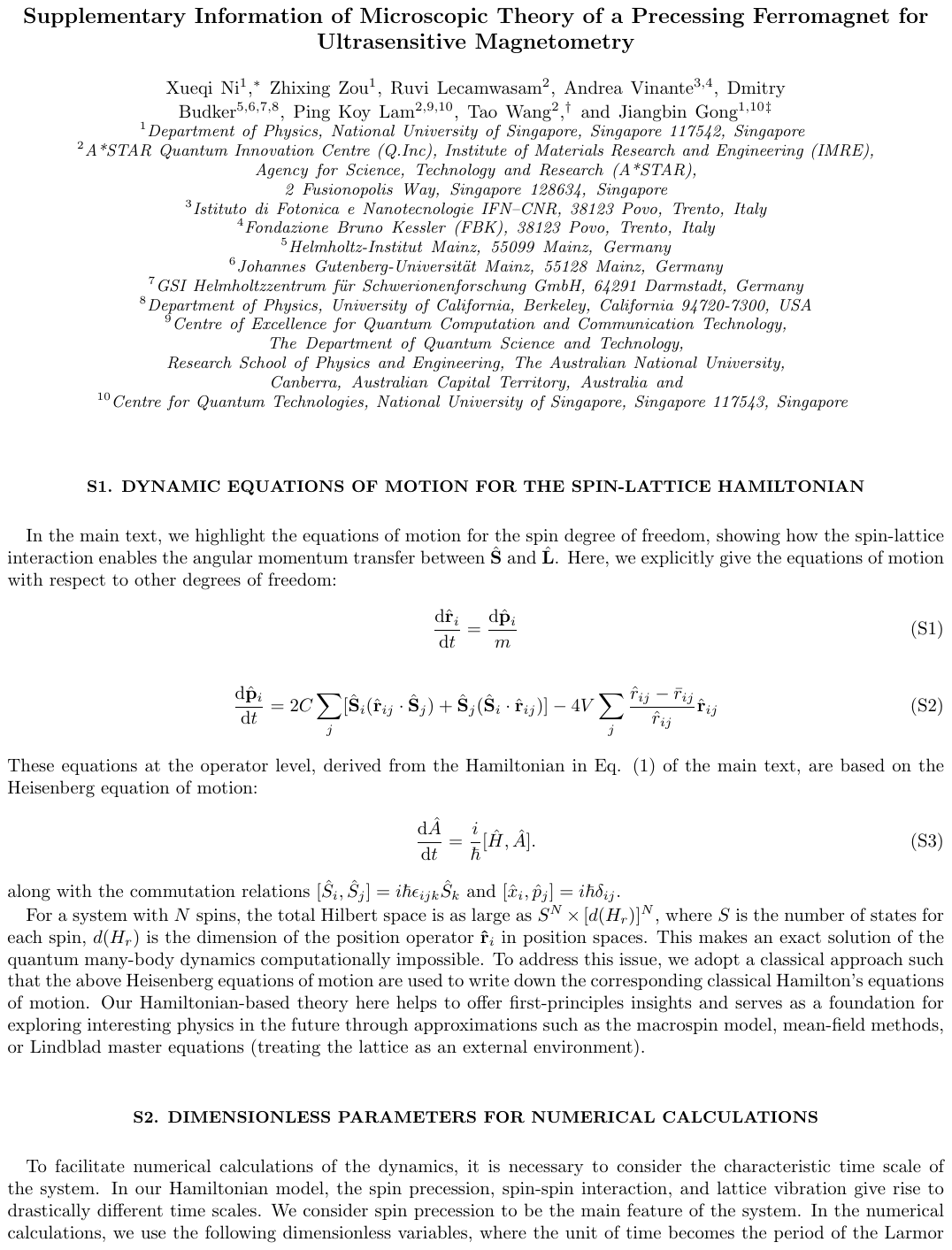}

\pdfximage{\supplementfilename}
\def\numbersupplementpages{\the\pdflastximagepages}

\newif\ifarXiv
\arXivtrue 
\usepackage{graphicx}
\usepackage{dcolumn}
\usepackage{bm}
\usepackage{color}
\usepackage{import}
\usepackage{hyperref}
\newcommand{\GJ}{\textcolor{black}}

\begin{document}

\preprint{APS/123-QED}

\title{Microscopic theory of a precessing ferromagnet for ultrasensitive magnetometry}

\author{Xueqi Ni$^1$}
\email{xueqi.ni@u.nus.edu}

\author{Zhixing Zou$^1$}%
\author{Ruvi Lecamwasam$^2$}%
\author{Andrea Vinante$^{3,4}$}
\author{Dmitry Budker$^{5,6,7,8}$}
\author{Ping Koy Lam$^{2,9,10}$}
\author{Tao Wang$^2$}%
\email{tao\_wang@imre.a-star.edu.sg}
\author{Jiangbin Gong$^{1,10}$}%
\email{phygj@nus.edu.sg}
\affiliation{
$^1$Department of Physics, National University of Singapore, Singapore 117542, Singapore
}

\affiliation{
$^2$A*STAR Quantum Innovation Centre (Q.Inc), Institute of Materials Research and Engineering (IMRE), Agency for Science, Technology and Research (A*STAR), 2 Fusionopolis Way, Singapore 128634, Singapore}

\affiliation{
$^3$Istituto di Fotonica e Nanotecnologie IFN–CNR, 38123 Povo, Trento, Italy}

\affiliation{
$^4$Fondazione Bruno Kessler (FBK), 38123 Povo, Trento, Italy}

\affiliation{
$^5$Helmholtz-Institut Mainz, 55099 Mainz, Germany}
\affiliation{
$^6$Johannes Gutenberg-Universit{\"a}t Mainz, 55128 Mainz, Germany}

\affiliation{
$^7$GSI Helmholtzzentrum f{\"u}r Schwerionenforschung GmbH, 64291 Darmstadt, Germany}

\affiliation{$^8$Department of Physics, University of California, Berkeley, California 94720-7300, USA}

\affiliation{
$^9$Centre of Excellence for Quantum Computation and Communication Technology, The Department of Quantum Science and Technology, Research School of Physics and Engineering, The Australian National University, Canberra, Australian Capital Territory, Australia}

\affiliation{
$^{10}$Centre for Quantum Technologies, National University of Singapore, Singapore 117543, Singapore
}

\begin{abstract}
\GJ{Levitated systems have great potential in quantum sensing and exploring quantum effects at the macroscopic scale. Of particular interest are recent works suggesting that a levitated ferromagnet can beat the standard quantum limit of magnetometry.  This work offers a {theoretical model} to analyze and understand critical features of the precessing dynamics of a levitated ferromagnetic needle, indeed much like a macrospin, in the presence of a weak magnetic field. The dynamics {from the atomic scale} reveals how the standard quantum limit is surpassed, thus verifying sensing advantages when compared with a collection of independent spins.  Our theory further takes us to two additional experimental designs of immediate interest: measurement of the celebrated Berry phase with a precessing ferromagnetic needle and the use of its nutation motion to sense a low-frequency oscillating magnetic field.  With a microscopic theory established for levitated ferromagnetic needles, future studies of macroscopic quantum effects and the associated quantum-classical transition also become possible.}

\end{abstract}

\maketitle

{\it Introduction}. Observing, controlling, and utilizing quantum mechanics phenomena on \GJ{scalable} platforms is a key direction for future technology development. Quantum sensing is one of the most mature quantum technologies, integrating fundamental physics and advanced applications\,\cite{RevModPhys.89.035002,bongs2023quantum,jiang2024searches,bass2024quantum,aslam2023quantum}.  Relying on the response of a quantum system to external parameters, \GJ{quantum sensing scales up its precision and sensitivity as we boost the measurement time $t$ and the number of particles involved (denoted $N$)}. For example, spin-exchange relaxation-free (SERF) magnetometers based on alkali-metal atomic vapors can achieve magnetic field sensitivity down to $0.16~\mathrm{fT/\sqrt{Hz}}$\,\cite{dang2010ultrahigh,opticalm}. Yet, for independent and identically distributed (IID) particles, the uncertainty in frequency estimation is bounded by the standard quantum limit (SQL)\,\cite{PhysRevLett.79.3865,huang_entanglement-enhanced_2024}, given by $\Delta \omega \geq 1/\sqrt{NT_2 t}$, where $T_2$ is the characteristic coherence time. Two strategies to enhance sensing sensitivity for a fixed $N$ are to extend $T_2$ or to utilize quantum entanglement\,\cite{doi:10.1126/science.1104149}. Along these directions, current state-of-the-art experiments can obtain a coherence time of several minutes \,\cite{yang2024minute,PhysRevLett.133.133202,young2020half,panda2024coherence}, and researchers have indeed demonstrated the use of maximally entangled states to surpass the SQL\,\cite{franke2023quantum,doi:10.1126/science.abi5226}. \GJ{Despite these breakthroughs, key challenges remain to advance quantum sensing further. In particular, as $N$ and hence the system size increases, the complexity of generating entangled states in a controlled manner grows. Furthermore, larger systems become more susceptible to decoherence and in a broad class of situations, decoherence suppresses sensing advantages arising from entanglements\,\cite{PhysRevLett.93.173002}.  These important issues in quantum sensing are therefore closely related to studies of macroscopic quantum phenomena\,\cite{stickler2018probing,PhysRevLett.110.160403,PhysRevLett.91.130401,PhysRevLett.132.023601,PhysRevA.105.012824,doi:10.1126/science.adf7553} and the quantum-to-classical transition in massive systems.}

\GJ{The levitated ferromagnetic needle, proposed by Kimball {\it et al.} in 2016\,\cite{PhysRevLett.116.190801}, emerged as an innovative sensing platform capable of achieving magnetic field sensitivity beyond the SQL without requiring quantum entanglement.} 
Experiments demonstrating the superconducting levitation of micron-scale ferromagnetic particles\,\cite{wang2019dynamics} and reporting magnetometry surpassing the energy resolution limit\,\cite{ahrens2024levitated,RevModPhys.92.021001, PhysRevLett.127.070801,PhysRevLett.133.263201} confirm that a levitated ferromagnetic needle platform is becoming experimentally realizable. Although Kimball \textit{et al.} offered stimulating physical insights into why the platform can beat the SQL, {a theory using only atomic-scale interactions to explain and verify the underlying physics behind ultrasensitive magnetometry at mesoscopic or macroscopic scales} is still lacking.

\GJ{Based on an explicit spin-lattice interaction term, we use a microscopic Hamiltonian to computationally investigate the sensing potential of a levitated ferromagnetic needle. This is an important step to (i) guide future material designs of ferromagnet-based quantum sensing, (ii) tune system parameters to analyze their respective roles, (iii) propose new applications of the levitated ferromagnet platform, and (iv) motivate future adventures in the creation of macroscopic quantum superpositions of a precessing needle.  The main challenge in the microscopical modeling of a levitated ferromagnetic needle is that we must treat many degrees of freedom with drastically different time scales (namely, those of spin precession, lattice vibrations, spin-lattice coupling, and spin-spin interaction, etc).  Notably, as we shall show below, even a classical treatment of an explicit spin-lattice interaction suffices to account for all the main features of needle magnetometry with predictive power\,\cite{BELOVS2025172735}. Our microscopic theory explains how all the spins in the lattice are effectively locked as one gigantic spin in coordination with the needle's macroscopic dynamics. With the role of other system parameters clarified,  this work lays a solid foundation for levitated ferromagnet-based magnetometry. More importantly,  we advance this research direction by proposing two experimental designs assisted by our computational simulations: the measurement of the celebrated Berry phase using a precessing ferromagnetic needle and the detection of a weak oscillating magnetic field\,\cite{PhysRevD.110.115029} using the needle nutation motion. }

\begin{figure}[h]
    \centering  \includegraphics[width=0.48\textwidth]{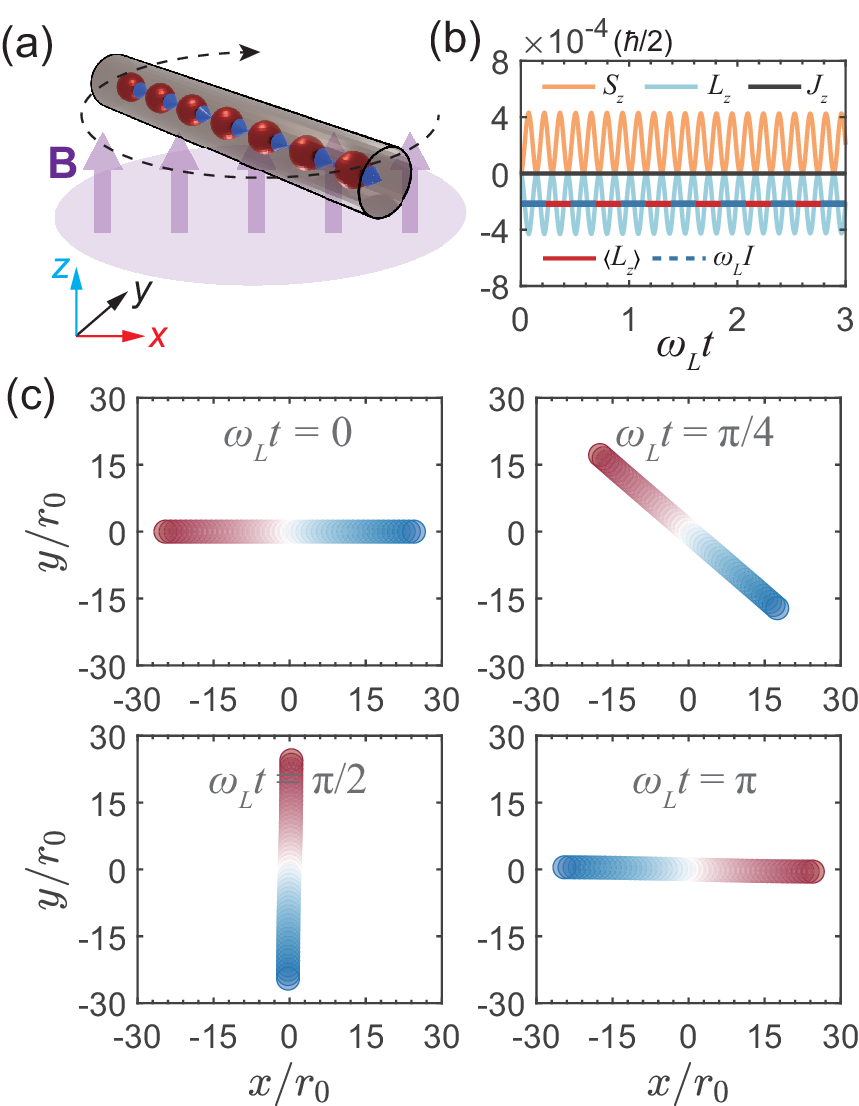}
    \caption{The precession dynamics of a levitated ferromagnetic needle, as predicted by the microscopic Hamiltonian in Eq.~(1).  (a) Schematic of a levitated macroscopic ferromagnetic needle over a superconductor. (b) Einstein-de Haas effect shows the transfer between spin angular momentum $S_z$ and mechanical angular momentum $L_z$. (c) The configuration of an atomic chain with $N$ = 50 atoms at different times. $r_0$ is the lattice constant. For cobalt, $r_0$ is 250.71 pm. Colors denote atoms with different initial positions.}
    \label{fig1}
\end{figure}

\GJ{{\it Microscopic modeling of needle precession dynamics.} We model a levitated ferromagnetic needle by a one-dimensional (1D) atomic lattice. In addition to the familiar spin-spin exchange interaction of strength $J$, one crucial physical term we identified from the literature is the pseudo-dipolar interaction\,\cite{assmann2019spin,strungaru2021spin}, whose strength is denoted by $C$.  With such an explicit interaction to couple the spins to the lattice orientation, we are able to write down the following total Hamiltonian as our microscopic theory:
\begin{equation}
\begin{aligned}
    H & = -\sum_i \gamma \mathbf{B}\cdot \mathbf{\hat{S}}_i -J\sum_{i}\mathbf{\hat{S}}_i \cdot \mathbf{\hat{S}}_{i+1} \\ & -C \sum_{ij} (\mathbf{\hat{S}}_i\cdot \mathbf{\hat{r}}_{ij})(\mathbf{\hat{r}}_{ij} \cdot \mathbf{\hat{S}}_j) + \sum_i \frac{\mathbf{\hat{p}}_i^2}{2m} + \sum_{ij}V (\hat{r}_{ij} - \bar{r}_{ij})^2,
\end{aligned}  
\end{equation}
where $\gamma \approx 1.76\times 10^{11}~\mathrm{rad\cdot s^{-1}\cdot T^{-1}}$ is electron gyromagnetic ratio, $m$ is the mass of the atom, ${\bf B}$ is an external magnetic field to be detected and measured, $\mathbf{\hat{r}}_{ij}$ denotes the displacement vector between two nearest neighboring atoms $i$ and $j$, and $\bar{r}_{ij}$ is the corresponding equilibrium distance. The lattice vibrations are modeled by a harmonic potential with a coupling strength $V$.  Due to the levitation, there is no additional external force or other interaction with a substrate. Below we use dimensionless variables where the time $t$ is rescaled by the characteristic Larmor frequency $\omega_L = \gamma \lvert \mathbf{B}\lvert$ and the spin is rescaled by the largest eigenvalue of $z$-component $\mathbf{s}_i = \mathbf{S}_i/S_0$. As a result, the spin-lattice coupling constant is mapped to $C_0 = (S_0^2/m\omega_L^2)C$ (Supplementary Material). Further, each spin is assumed to be spin-1/2 and hence $S_0=\hbar/2$ while it can be readily extended to larger spin systems.} 


\GJ{Exactly solving the quantum many-body dynamics governed by the Hamiltonian in Eq.~(1) is practically impossible.  Fortunately, as confirmed by our classical trajectory simulations below, all the spins can stay ``coherent" with each other, indicating essentially zero entanglement between the spins or between the spins and the lattice motion \footnote{To appreciate how classical simulations help to assess entanglement generation, please see\,\cite{gong1,gong2}}.  As such,  we resort to fully classical equations of motion to investigate the underlying physics, replacing
$\hat{\mathbf{S}}$, $\hat{\mathbf{r}}$, and $\hat{\mathbf{p}}$ with three-dimensional classical vectors.  In particular, the equation of motion in terms of the spins is given by:
\begin{eqnarray}
\begin{aligned}
    \frac{\mathrm{d} \mathbf{S}_i}{\mathrm{d} t} & = \gamma\mathbf{S}_i \times \mathbf{B} + J \mathbf{S}_i\times ( \mathbf{S}_{i+1} +\mathbf{S}_{i-1}) \\ & +2C\mathbf{S}_i\times \sum_j \mathbf{r}_{ij}(\mathbf{r}_{ij} \cdot \mathbf{S}_j).
\end{aligned}  
\end{eqnarray}
The equations of motion for the lattice coordinate $\mathbf{r}_i$ and momentum $\mathbf{p}_i$ are given in the Supplementary Material.
Note that the time scale of lattice vibration differs from that of the characteristic collective needle dynamics presented below by around 10 orders of magnitude. Though this huge time scale mismatch justifies an option to freeze the lattice vibrational motion altogether, we choose to use a softer lattice potential to facilitate our dynamics simulations and better understand the spin-lattice coupling dynamics.  Likewise, the ferromagnetic spin-spin interaction also leads to a time scale many orders of magnitude smaller than the needle dynamics.  We shall artificially tune the spin-spin interaction strength $J$ to understand its main physical role.  For example, in Supplementary Material, 
we present how to extract the phenomenological Landau-Lifshitz-Gilbert (LLG) damping from our microscopic modeling, concluding that the actual strength $J$ has a negligible effect on the LLG damping coefficient. 
}
 
\GJ{Figure 1 shows computational results based on Hamiltonian dynamics governed by the equations of motion mentioned above. 
Simulations shown in Fig.~1 are for a 1D atomic chain comprising 50 cobalt atoms.  
Initially, all the atoms are arranged along the $x$-axis according to the lattice constant of cobalt. An external magnetic field is applied along the $z$-direction with an amplitude of $1~\mathrm{nT}$, as illustrated in Fig. 1(a). The exchange interaction energy $J$ is chosen to be $10^4$ times the Zeeman interaction energy, thus offering a strong enough intrinsic magnetic field to protect the ferromagnetic phase. The choice of the spin-lattice constant $C_0$ is 1.2$\times 10^5$ to match the experimental Gilbert damping coefficient (Supplementary Material), but the actual value is not essential to the results presented here.  Figure 1(b) verifies the Einstein–de Haas effect\,\cite{dornes2019ultrafast,PhysRevLett.112.085503,PhysRevLett.129.257201,NIE2025301}, where $S_z$, the total spin angular momentum along $z$-axis, and $L_z$, the lattice mechanical angular momentum along $z$-axis, keep exchanging, but with their sum conserved over time.  That is, during the dynamical evolution, $J_z = S_z + L_z$ is a conserved quantity. Remarkably, the ferromagnetic needle as a whole is seen to acquire a mechanical angular momentum and start to precess collectively around the $z$ axis.  This precession behavior is much like that of a single spin initially polarized in the $x$-$y$ plane.  If there were one single spin, the applied magnetic field would lead to the Larmor frequency $\omega_L$. For a 1~nT magnetic field, the frequency is 28~Hz. Now with spin-lattice coupling, the spins are seen to drive the entire lattice to precess around the $z$-axis. Indeed, Fig.~1(c) shows the explicit configuration of the 1D atomic chain, obtained from the dynamics simulations, at 0, 1/8, 1/4, and 1/2 of the precession period $2\pi/\omega_L$.   A continuous animation is also attached in the Supplementary Media. Interestingly, the small oscillation in $L_z$ (as depicted in Fig.~1b) does indicate that the following of the needle orientation with the spin precession is not instantaneous -- only upon averaging out the fast oscillations, $L_z$ depicts a perfect precession with the same frequency $\omega_L$. It is important to note that in our simulations, there is no need to fine-tune the initial conditions.} For instance, we can randomly sample different initial configurations of the lattice according to the Boltzmann distribution at a given temperature $T_L$ (Supplementary Material). Though the microscopic variables vary for different individual trajectories, the macroscopic collective motion as depicted in 
Fig.~1 persists.


\begin{figure}[h]
    \centering   \includegraphics[width=0.48\textwidth]{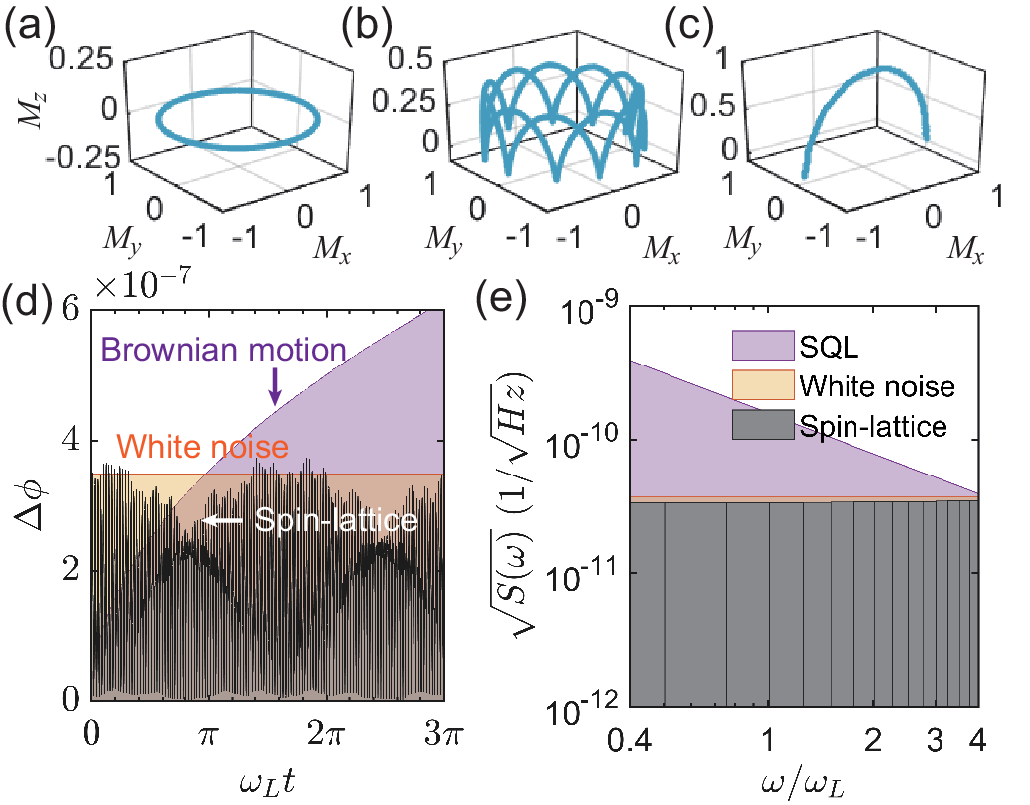}
    \caption{(a)-(c) Three dominating dynamical modes of needle collective motion including (a) precession, (b) nutation, and (c) libration, over two periods $4\pi/\omega_L$.  \GJ{(d) Standard deviations in the precession angle $\phi$ as a function of time, under different scenarios. (e) Bar plot of noise power spectral density due to spin-lattice interaction, as compared with that of white noise (which is similar) and that in the case of a collection of independent spins illustrating the SQL.}}
    \label{fig2}
\end{figure}

\GJ{{\it Different modes of needle collective motion.} The precession motion seen above is modified at stronger magnetic fields. Indeed, assuming pure precession dynamics,  we have $|S_z| = |L_z| \approx \omega_L I$, where $I$ is the moment of inertia of the needle, with $I \approx \frac{1}{12}mr_0^2 N^3$. Because the spin angular momentum $S_z$ is at the order of $N\hbar/2$, for collective lattice precession to dominate we must have $\omega_L I \ll N\hbar/2$, corresponding to a magnetic field of $B\ll B_{\rm c} \approx 6\hbar/(\gamma m r_0^2N^2)$.}  For instance, $B_{\rm c}$ is approximately $ 230~\mathrm{\mu T}$ given 50 cobalt atoms. 
\GJ{It is then interesting to observe, from our first-principle simulations, the dynamics of the needle if the magnetic field strength increases.  As shown in Figs.~2(a)-(c), the collective motional modes of the needle gradually change from precession to nutation and eventually to libration around the magnetic field direction. Specifically, Figs.~2(a)-(c) depict the time evolution of the normalized magnetization $\mathbf{M}$, defined as $\frac{1}{N}\sum_i \mathbf{s_i}$, in three-dimensional space under magnetic fields of $1~\mathrm{nT}$, $50~\mathrm{\mu T}$, and $5~\mathrm{mT}$, respectively. Under weak magnetic fields, the collective precession of the lattice dominates, and the trajectory of the magnetization forms a unit circle in the $x$-$y$ plane over time (Fig.~2a). As the magnetic field strength increases, nutation becomes increasingly significant (Fig.~2b). When the magnetic field exceeds $B_{\rm c}$, all spins tend to align parallel to the field direction (Fig.~2c). In this regime, the magnetization oscillates in the $z$-$x$ plane, where $z$ ($x$) is the direction of the magnetic field (initial lattice orientation). Over a long time of evolution, the plane of the libration will also rotate at an intrinsic frequency (Supplementary Material). Using a microscopic theory, our results confirm and strengthen an earlier study that predicts the three different dynamical regimes with a macrospin model\,\cite{PhysRevLett.121.160801}.  
Significantly, in all three dynamical regimes, we always have $M_x^2+M_y^2+M_z^2=1$, indicating that the macrospin $(M_x, M_y, M_z)$ stays on the Bloch sphere, and there is no observable dephasing. That is, all individual spins (upon fast averaging around the needle axis) have the same polarization. As mentioned earlier, this is one strong justification for a classical treatment of the spin degree of freedom.}





\GJ{{\it Beating the SQL: a microscopic dynamics perspective.}  Next we computationally investigate how a levitated ferromagnetic needle can sense a magnetic field with a precision surpassing the SQL.  To that end, we first elaborate on the SQL using an ensemble of independent spins to sense a magnetic field $\mathbf{B}$.  The noise independently experienced by the spins necessarily causes each spin's precession motion to fluctuate with time.  This process can be modeled by a stochastic noise term so that the total energy is given by $\sum_{i} (\mathbf{B}+\mathbf{\xi}_i) \cdot \mathbf{S}_i$, where $\mathbf{\xi}_i$ represents the independent white noise experienced by each spin.  In this picture,  the ensemble of the spins undergoes a random walk. Let $\Delta \phi = \sqrt{\frac{1}{N} \sum_i (\phi_i - \langle \phi \rangle)^2}$ be the uncertainty in precession angle $\phi$. Then it grows as $\Delta \phi \propto \sqrt{t}$ (Fig.~2d) and the corresponding noise power spectral density (PSD) $S(\omega)$ has a typical $1/\omega^2$ scaling (Fig.~2e)}\,\cite{krapf2018power} (Supplementary Material).  \GJ{Consequently, quantum sensing using independent spins is constrained by the SQL, where uncertainty in the precession frequency estimation follows $\Delta \omega = \Delta \phi/t \propto 1/\sqrt{t}$. }

\GJ{By contrast, for a levitated ferromagnetic needle, despite the noise due to the intrinsic spin-lattice coupling,  the uncertainty $\Delta \phi$ during the precession dynamics saturates instead of the characteristic diffusive behavior.  Microscopically, this verifies the qualitative insight by Kimball {\it et al.}\,\cite{PhysRevLett.116.190801}.
Due to the saturation of the precession angle uncertainty, all the spins remain ``in phase", thus explaining the observation above that the macrospin does stay on the Bloch sphere. Indeed, the spin-spin exchange interaction and the spin-lattice interaction jointly lead to a neat dynamical decoupling effect\,\cite{viola1998dynamical,viola1999dynamical}: any noise causing the spins to drift away from the lattice axis is canceled by fast rotation around the lattice axis. Interestingly, microscopic dynamics here also allows us to track the explicit time dependence of the uncertainty in $\Delta \phi$: for the parameters chosen in the shown simulations, it exhibits oscillations more than 2 orders of magnitude faster than the precession frequency itself. The rate of this important self-averaging is found to be directly determined by the strength of the spin-lattice coupling.   We further examine the PSD of the intrinsic noise due to spin-lattice coupling and spin-spin interaction. It is seen to have a characteristic flat spectrum (Fig.~2e), hence close to that of white noise.   Using the white-noise limit and the associated Cramér-Rao lower bound (CRLB), the scaling of uncertainty in frequency estimation is given by $\Delta \omega \geq \sqrt{\frac{12}{\mathrm{SNR}f_{\rm BW}t^3}}$,
where SNR means signal-to-noise ratio and $f_{\rm BW}= 1/\mathrm{d}t$ is the frequency bandwidth\,\cite{kay_fundamentals_1993}. As the measurement time increases, $\Delta \omega$ scales as $t^{-3/2}$, thus beating the SQL. Unlike the realization of cooperative spins using feedback control\,\cite{PhysRevLett.133.133202}, here the spin-spin correlation inherent in a ferromagnetic needle offers an intrinsic approach to surpassing the SQL}.

\begin{figure}[ht]
    \centering   \includegraphics[width=0.48\textwidth]{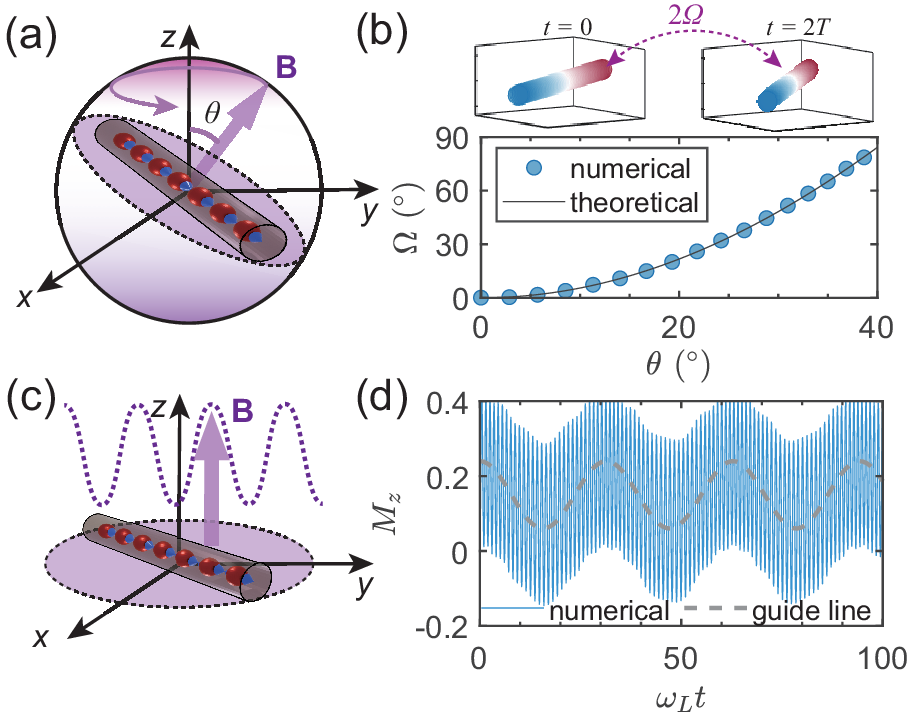}
    \caption{\GJ{(a) Schematic of a ferromagnetic needle in the presence of a slowly rotating magnetic field with a fixed polar angle $\theta$.  (b) The initial and final configurations of the needle differ by an angle $2\Omega = 4\pi(1-\cos\theta)$, manifesting the Berry phase associated with the adiabatic protocol.  (c) Schematic of a needle undergoing nutation in an oscillating magnetic field. (d) Results from the microscopic dynamics, indicating that the oscillation in the magnetic field is mapped to the time dependence of the nutation amplitude}.}
    \label{fig3}
\end{figure}


\GJ{{\it Berry Phase manifested by a levitated needle.}  Armed with our microscopic theory, we next propose to detect the celebrated Berry phase using the precession dynamics of a levitated ferromagnetic needle. This will be of fundamental interest to our understanding of quantum-classical correspondence and to quantum metrology.  We first describe the protocol using a quantum mechanics language for a collection of $N$ isolated spins.  If all the spins are initially polarized perpendicular to the external field, the total spin state is a product state, i.e.,  $ [\frac{1}{\sqrt{2}} (\lvert\uparrow\rangle + \lvert\downarrow\rangle )]^{\otimes N}$,  where $\lvert\uparrow\rangle$ and $\lvert\downarrow\rangle$ are spin-up and spin-down states with respect to the field direction. If the applied magnetic field $\mathbf{B}$ rotates slowly (as compared with the precession frequency) around the $z$-axis (Fig.~3a) and completes one cycle, the two spin components $\lvert\uparrow\rangle$ and $\lvert\downarrow\rangle$ accumulate Berry phases $\gamma^{B}_{\uparrow}$  and $\gamma^{B}_{\downarrow}=-\gamma^{B}_{\uparrow}$ respectively, on top of their dynamical phases.  Similar to how experiments in NMR, NV center, and atomic magnetometer systems\,\cite{suter1987berry,arai2018geometric,wang2025pulsed} extracted the Berry phase, in the second cycle we repeat the same process but reverse the applied magnetic field, thus canceling the dynamical phases.
This two-cycle protocol then leads to the final spin state $[\frac{1}{\sqrt{2}} (e^{2i\gamma^{B}_{\uparrow}}\lvert\uparrow\rangle + e^{2i\gamma^{B}_{\downarrow}}\lvert\downarrow\rangle )]^{\otimes N}$, a state evidently misaligned from the initial orientation by $2\Omega=2\gamma^{B}_{\uparrow} - 2\gamma^{B}_{\downarrow}$, where $\Omega$ is the solid angle traced out by the rotating magnetic field.  Our dynamical simulation shows that this misalignment arising from a geometrical effect is indeed passed to the final configuration of the needle.  In particular, we assume that $\mathbf{B}(t)$ is rotating around the $z$-axis at a fixed polar angle $\theta$.  As shown in Fig.~3b, upon reading out the final orientation of the needle, the misalignment from its initial configuration can be found and compared with the theoretical result $2\Omega$, with $\Omega = 2\pi (1-\cos\theta)$. The results retrieved from our microscopic dynamics simulations of the needle are in excellent agreement with the theoretical Berry phase results. Berry phase enables stroboscopic measurements, enhancing robustness against external noise and quantum backaction\,\cite{PhysRevLett.106.143601}. Furthermore, we again demonstrate from microscopic dynamics that a ferromagnetic needle can behave like a macrospin, even in terms of highly subtle geometrical aspects of the spin dynamics.}

\GJ{{\it Frequency sensing of an oscillating magnetic field.} 
Finally, exploiting our microscopic theory, we propose to use the dynamical regime of nutation to determine the frequency of a magnetic field with a slowly oscillating amplitude, as illustrated in Fig.~3(c) and 3(d).  In the presence of an oscillating field, the precession, manifested by sinusoidal $M_x$ or $M_y$, becomes chirped due to its frequency dependence on the field amplitude, thus making the signal analysis more challenging.  Interestingly, nutation motion offers a simpler alternative for sensing because its amplitude $M_z$ is found to be directly proportional to the amplitude of the oscillating field. As an example, the nutation dynamics under an oscillating magnetic field $B_z = B_0 [1+0.5\cos(0.2\omega_Lt)]$ is shown in Fig.~3(d). There the blue line represents numerical results from the dynamics of the needle, whereas the dashed gray line serves as a guide of our eyes to show that the temporal profile of $B_z$ and $M_z$ coalesce. Using the Fourier analysis, we can then determine the oscillation frequency of the applied magnetic field.  Already supported by first-principle simulations here, the nutation motion of a levitated needle is expected to be useful for the sensing of axion-like dark matter\,\cite{gramolin2021search,jiang2021search,PhysRevD.110.115029} that is often connected with an oscillating field.  Using some analysis similar to the case of precession dynamics, it is straightforward to find that here the precision limit in determining the oscillation frequency of the field also surpasses the SQL.} More details of this sensing protocol are elaborated in Supplementary Material.

{\it Conclusion.} \GJ{With an explicit spin-lattice interaction accounted for, we are able to fully investigate the {atomic-scale} dynamics of a levitated ferromagnetic needle as a sensing platform.   The phenomenological damping due to spin-lattice coupling, including its quantum analog\,\cite{PhysRevLett.133.266704,PhysRevLett.110.147201,anders2022quantum,uhrig2025landaulifshitzdampinglindbladiandissipation}, can now be investigated from a Hamiltonian theory. Both conceptually and computationally, we have demonstrated the great potential of the levitated ferromagnet platform in ultrasensitive magnetometry. The demonstration of how the Berry phase can be measured by a precessing needle shall motivate future studies, especially on new possibilities in connecting single-spin dynamical behaviors with that of a macroscopic ferromagnetic needle.  Our microscopic theory thus offers a powerful toolbox to guide future experimental studies. For example, we advocate using our model to investigate possible backaction effects in actual sensing measurements and to study the possible creation of a macroscopic quantum superposition of the rotational states of a ferromagnetic needle. }

{\it Acknowledgements.}  The authors thank Prof. Alex Sushkov for valuable discussions. J.G., P.K.L., and T.W. acknowledge the support of the National Research Foundation, Singapore, under its Competitive Research Programme (CRP Award No: NRF-CRP30-2023-0002). Any opinions, findings, conclusions, or recommendations expressed in this material are those of the author(s) and do not reflect the views of the National Research Foundation, Singapore. T.W. acknowledges support from Italy-Singapore science and technology collaboration grant (R23I0IR042), Delta-Q (C230917004, Quantum Sensing). A.V. acknowledges financial support from the Italian Ministry for University and Research within the Italy-Singapore Scientific and Technological Cooperation Agreement 2023-2025. A.V. and D.B. acknowledge support from the QuantERA II Programme (project LEMAQUME) that has received funding from the European Union’s Horizon 2020 research and innovation programme under Grant Agreement No. 101017733.



\bibliography{microscopic_sl}%

\ifarXiv
    

\section*{Supplementary material}

# Supplementary Information of Microscopic Theory of a Precessing Ferromagnet for Ultrasensitive Magnetometry

Xueqi Ni[1],* Zhixing Zou[1], Ruvi Lecamwasam[2], Andrea Vinante[3,4], Dmitry Budker[5,6,7,8], Ping Koy Lam[2,9,10], Tao Wang[2],† and Jiangbin Gong[1,10‡]

[1]*Department of Physics, National University of Singapore, Singapore 117542, Singapore*
[2]*A*STAR Quantum Innovation Centre (Q.Inc), Institute of Materials Research and Engineering (IMRE),
Agency for Science, Technology and Research (A*STAR),
2 Fusionopolis Way, Singapore 128634, Singapore*
[3]*Istituto di Fotonica e Nanotecnologie IFN–CNR, 38123 Povo, Trento, Italy*
[4]*Fondazione Bruno Kessler (FBK), 38123 Povo, Trento, Italy*
[5]*Helmholtz-Institut Mainz, 55099 Mainz, Germany*
[6]*Johannes Gutenberg-Universität Mainz, 55128 Mainz, Germany*
[7]*GSI Helmholtzzentrum für Schwerionenforschung GmbH, 64291 Darmstadt, Germany*
[8]*Department of Physics, University of California, Berkeley, California 94720-7300, USA*
[9]*Centre of Excellence for Quantum Computation and Communication Technology,
The Department of Quantum Science and Technology,
Research School of Physics and Engineering, The Australian National University,
Canberra, Australian Capital Territory, Australia and*
[10]*Centre for Quantum Technologies, National University of Singapore, Singapore 117543, Singapore*

## S1. DYNAMIC EQUATIONS OF MOTION FOR THE SPIN-LATTICE HAMILTONIAN

In the main text, we highlight the equations of motion for the spin degree of freedom, showing how the spin-lattice interaction enables the angular momentum transfer between $\hat{\mathbf{S}}$ and $\hat{\mathbf{L}}$. Here, we explicitly give the equations of motion with respect to other degrees of freedom:

$$\frac{d\hat{\mathbf{r}}_i}{dt} = \frac{d\hat{\mathbf{p}}_i}{m} \tag{S1}$$

$$\frac{d\hat{\mathbf{p}}_i}{dt} = 2C \sum_j [\hat{\mathbf{S}}_i (\hat{\mathbf{r}}_{ij} \cdot \hat{\mathbf{S}}_j) + \hat{\mathbf{S}}_j (\hat{\mathbf{S}}_i \cdot \hat{\mathbf{r}}_{ij})] - 4V \sum_j \frac{\hat{r}_{ij} - \bar{r}_{ij}}{\hat{r}_{ij}} \hat{\mathbf{r}}_{ij} \tag{S2}$$

These equations at the operator level, derived from the Hamiltonian in Eq. (1) of the main text, are based on the Heisenberg equation of motion:

$$\frac{d\hat{A}}{dt} = \frac{i}{\hbar}[\hat{H}, \hat{A}]. \tag{S3}$$

along with the commutation relations $[\hat{S}_i, \hat{S}_j] = i\hbar \epsilon_{ijk} \hat{S}_k$ and $[\hat{x}_i, \hat{p}_j] = i\hbar \delta_{ij}$.

For a system with $N$ spins, the total Hilbert space is as large as $S^N \times [d(H_r)]^N$, where $S$ is the number of states for each spin, $d(H_r)$ is the dimension of the position operator $\hat{\mathbf{r}}_i$ in position spaces. This makes an exact solution of the quantum many-body dynamics computationally impossible. To address this issue, we adopt a classical approach such that the above Heisenberg equations of motion are used to write down the corresponding classical Hamilton's equations of motion. Our Hamiltonian-based theory here helps to offer first-principles insights and serves as a foundation for exploring interesting physics in the future through approximations such as the macrospin model, mean-field methods, or Lindblad master equations (treating the lattice as an external environment).

## S2. DIMENSIONLESS PARAMETERS FOR NUMERICAL CALCULATIONS

To facilitate numerical calculations of the dynamics, it is necessary to consider the characteristic time scale of the system. In our Hamiltonian model, the spin precession, spin-spin interaction, and lattice vibration give rise to drastically different time scales. We consider spin precession to be the main feature of the system. In the numerical calculations, we use the following dimensionless variables, where the unit of time becomes the period of the Larmor

precession, as listed in Table. 1.

| Physical quantity | Dimensionless variable |
|---|---|
| time $t$ | $t_0 = \omega_L t$ |
| spin number $\mathbf{S}$ | $\mathbf{s} = \mathbf{S}/S_0$ |
| magnetic field $\mathbf{B}$ | $\mathbf{b} = \mathbf{B}/|\mathbf{B}|$ |
| position $\mathbf{r}$ | $\mathbf{r}' = \sqrt{\frac{\gamma|\mathbf{B}|m}{S_0}}\mathbf{r}$ |
| momentum $\mathbf{p}$ | $\mathbf{p}' = \sqrt{\frac{1}{\gamma|\mathbf{B}|S_0 m}}\mathbf{p}$ |
| spin-spin interaction $J$ | $J_0 = \frac{S_0}{\gamma|\mathbf{B}|}J$ |
| spin-lattice coupling $C$ | $C_0 = \frac{S_0^2}{m\gamma^2|\mathbf{B}|^2}C$ |
| Harmonic potential $V$ | $V_0 = \frac{1}{m\gamma^2|\mathbf{B}|^2}V$ |

Table I: Dimensionless variables used in the calculations.

By using the dimensionless variables, the dynamics equations of motion become:

$$\frac{d\mathbf{s}_i}{dt_0} = \mathbf{s}_i \times \mathbf{b} + J_0 \mathbf{s}_i \times (\mathbf{s}_{i+1} + \mathbf{s}_{i-1}) + 2C_0 \mathbf{s}_i \times \sum_j \mathbf{r}'_{ij}(\mathbf{r}'_{ij} \cdot \mathbf{s}_j), \tag{S4}$$

$$\frac{d\mathbf{r}'_i}{dt_0} = \mathbf{p}'_i, \tag{S5}$$

$$\frac{d\mathbf{p}'_i}{dt_0} = 2C_0 \sum_j [\mathbf{s}_i(\mathbf{r}'_{ij} \cdot \mathbf{s}_j) + \mathbf{s}_j(\mathbf{s}_i \cdot \mathbf{r}'_{ij})] - 4V_0 \sum_j \frac{r'_{ij} - \bar{r}'_{ij}}{r'_{ij}} \mathbf{r}'_{ij}. \tag{S6}$$

We solve the equations of motion using the fourth-order Runge–Kutta (RK4) method with a time step of 0.0001 in MATLAB. The results are consistent with those obtained using higher-order methods (e.g., RK8) and stiff solvers like Rodas4P (a fourth-order A-stable Rosenbrock method with stiff-aware interpolation) implemented in Julia.

### S3. CONNECTION TO THE LANDAU-LIFSHITZ-GILBERT EQUATION

The pioneering proposal of needle magnetometry largely used the phenomenological Landau-Lifshitz-Gilbert (LLG) damping for physical reasoning. The LLG equation is widely used in micromagnetic simulations for spintronics applications, but it is fair to say that its explicit physical origin has not yet been fully understood [1,2]. To connect actual material parameters and physical properties of a system with LLG damping to facilitate experimental designs, we exploit our microscopic theory to dynamically estimate the LLG damping coefficient in our one-dimensional spin lattice model. In particular, the LLG equation is given by

$$\frac{d\mathbf{s}_i}{dt} = \gamma \mathbf{s}_i \times \mathbf{H}_{\text{eff}} + \gamma \eta \mathbf{s}_i \times (\mathbf{s}_i \times \mathbf{H}_{\text{eff}}), \tag{S7}$$

where $\mathbf{H}_{\text{eff}}$ is an effective magnetic field including spin-spin exchange interaction $\mathbf{H}_{\text{eff}} = \mathbf{B} + J/\gamma(\mathbf{S}_{i+1} + \mathbf{S}_{i-1})$, and $\eta$ is the Gilbert damping coefficient that quantifies the energy dissipation rate from the spins to the lattice. The first term $\gamma \mathbf{s}_i \times \mathbf{H}_{\text{eff}}$ results in a collective precession while the second term $\gamma \eta \mathbf{s}_i \times (\mathbf{s}_i \times \mathbf{H}_{\text{eff}})$ contribute to the damping. In terms of the microscopic dynamics derived from our model, the energy damping arises intrinsically from the pseudo-dipolar spin-lattice interaction $2C\mathbf{S}_i \times \sum_j \mathbf{r}_{ij}(\mathbf{r}_{ij} \cdot \mathbf{S}_j)$. Therefore, by comparing Eq. (S7) with the microscopic dynamics equations of motion as given by Eq. (2) in the main text, the effective damping coefficient in our model can be evaluated by performing a simple time average, namely,

$$\langle \eta \rangle = \frac{\langle 2C_0 \sum_j \mathbf{r}'_{ij}(\mathbf{r}'_{ij} \cdot \mathbf{s}_j) \rangle}{\langle \mathbf{s}_i \times \mathbf{H}_{\text{eff}}/|\mathbf{B}| \rangle}. \tag{S8}$$

From Eq. (S8), it is evident that $\langle \eta \rangle$ is proportional to the spin-lattice coupling strength $C_0$. In our simulations for a given magnetic field of 1 nT, we have set $C_0$ to $1.2 \times 10^5$ to match the experimental value of $\eta \approx 0.01$. The damping coefficient versus the spin-lattice coupling strength $C_0$ is shown in Fig. S1(a). Though spin exchange interaction contributes a strong internal magnetic field, it hardly changes the Gilbert damping because all the spins are essentially "in phase" in our simulation, meaning $\mathbf{S}_i \parallel \mathbf{S}_{i\pm 1}$ (Fig. S1(b)). Besides, our microscopic model incorporating the lattice vibration also allows us to explicitly study how the lattice harmonic potential strength $V_0$ influences the effective LLG damping coefficient. As Fig. S1(c) shows, the damping coefficient monotonously and slowly decreases as $V_0$ increases, at least until $V_0 = 10^8$ where we are forced to take a cut due to the huge time scale mismatch between the lattice vibration and the spin precession dynamics. In Fig. S1(d), we also investigate the temperature dependence of the effective Gilbert damping coefficient by changing the lattice temperature $T_L$ up to 0.5 mK. In levitated experiments, the temperature of nanoparticles has been cooled down to around 12 $\mu$K using feedback cooling[3]. Similarly, cooling down a levitated ferromagnetic needle would help suppress thermal noise and enable the exploration of quantum superposition. It should also be noted that the $t^{-3/2}$ sensitivity scaling remains unaffected at higher temperatures, provided the coherence between spins is preserved by inherent interactions. For pure ferromagnetic materials Fe, Co, and Ni, ferromagnetic resonance (FMR) measurements have observed that the Gilbert damping will increase as temperature decreases[4,5]. Our microscopic simulations do agree with this observed trend in the low-temperature regime accessible by our computational methods.

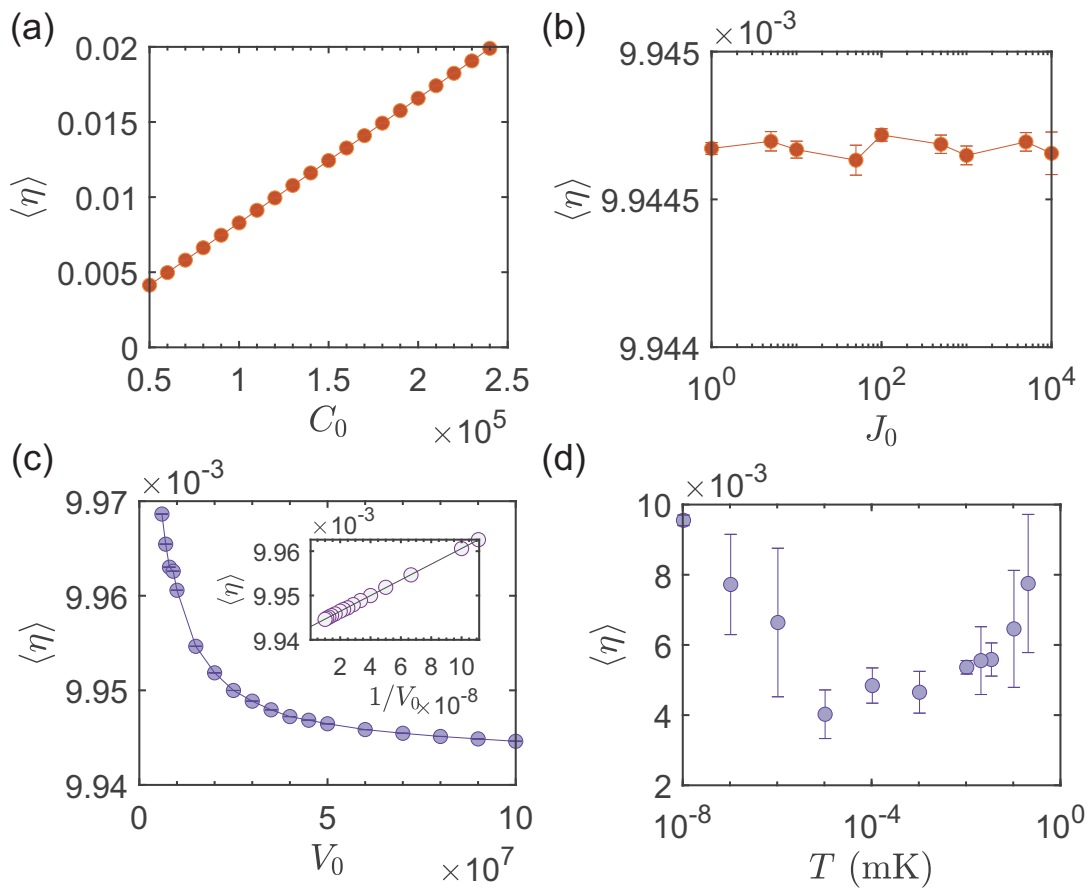

Figure S1: Effective Gilbert damping coefficients when varying (a) spin-lattice coupling strength $C_0$, (b) spin-spin exchange interaction strength $J_0$, (c) lattice vibration strength $V_0$, and (d) the effective temperature of the lattice.

## S4. RESULTS WITH VARYING NUMBERS OF ATOMS

The actual number of atoms involved in the calculation will not introduce qualitative differences from the results in the main text. In other words, there are no significant finite-size effects in our calculations. In this section, we

elaborate on this point. Firstly, we calculate the Gilbert damping coefficients similar to the last section, with varying numbers of atoms. It is shown in Fig. S2(a) that when $N$ changes by one order of magnitude (from 10 to 100), there is only a negligible change in the Gilbert damping coefficients. As such, our computer simulations confirm that Gilbert damping is a material property intrinsically determined by the coupling strengths $J$, $C$, $V$, and temperature $T$ and has little to do with the size of the material.

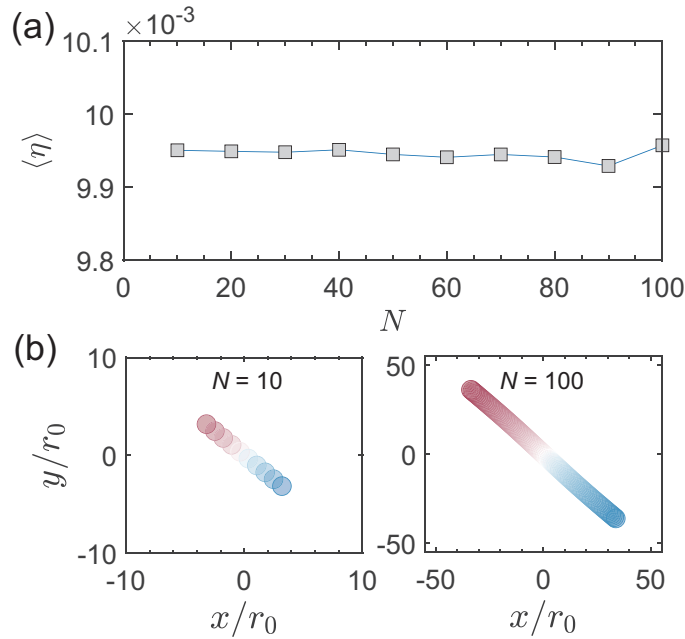

Figure S2: (a) Gilbert damping coefficient with varying numbers of atoms in the simulation. (b) Lattice configuration at $1/8$ of a precession period for $N = 10$ and $N = 100$ atoms.

Besides the Gilbert damping coefficients, in Fig. S2(b), we compare the explicit configuration of the lattice at the same time with $N = 10$ and $N = 100$ atoms under 1 nT magnetic field. They both present essentially perfect precession in the $x-y$ plane at the same precessing frequency. Therefore, though computational ability limits the total number of atoms that can be simulated, the main features are truly robust with respect to the number of atoms.

An important consideration due to an increase in the number of atoms is the maximum amplitude of the magnetic field to observe precession dominantly. As discussed in the main text, this value scales as $1/N^2$ for a one-dimensional ferromagnet. For $N = 10^6$ atoms, the corresponding magnetic field $B_c$ decreases to 580 fT. Experiments must therefore balance the size of the ferromagnetic material with the amplitude of the magnetic fields, as larger structures can only precess at lower frequencies. Fortunately, in real experiments we always have three-dimensional materials. For a given number of atoms $N$, $B_c$ will be much higher compared with the one-dimensional case illustrated in the main text. That is, because three-dimensional materials have a more compact structure, we expect to have a smaller moment of inertia as compared with the 1D case with the same number of atoms. Furthermore, the material's shape affects its moment of inertia and thus alters $B_c$. For a three-dimensional cylinder with radius $r$ and length $L$, the moment of inertia is given by $I = \frac{1}{12}\rho\pi r^2 L^3$ and the total number of spins is $N = \rho\pi r^2 L/m$, where $m$ is the mass of an atom and $\rho$ is the density. Comparing $\gamma B I$ and $N\hbar/2$, the corresponding magnetic field is now $B_c \approx 6\hbar/m\gamma L^2$. For a micro-scale needle, for instance $L = 1$ $\mu$m, $B_c$ is around 40 nT, a value achievable in experiments by proper magnetic shielding[6].

## S5. ROTATION OF THE LIBRATION PLANE

In this section, we provide more insight into the three different dynamical regimes. For the precession regime, the precession frequency of $M_x$ or $M_y$ is proportional to the external magnetic field $\omega_L = \gamma|\mathbf{B}|$. Therefore, we can use the precession motion to infer the amplitude of an external magnetic field. When the magnetic field increases, the transitions to nutation and libration are smooth. We can classify the three dynamical regimes, where each dynamic mode dominates, according to the maximum value of the total spin-$z$ component.

| | Precession | Nutation | Libration |
|---|---|---|---|
| max[$|S_z|$] | $\ll N\hbar/2$ | $< N\hbar/2$ | $= N\hbar/2$ |

Table II: Classification of three dynamical regimes.

For libration, which is the usual case of a heavier ferromagnet or higher external magnetic field, $L_z$ and $S_z$ saturate to $N\hbar/2$. Note that the plane of libration will still rotate over a long time of evolution since $L_z$ is non-zero. Fig. S3a shows the slight rotation of the libration plane over 10 periods $20\pi/\omega_L$. Here, $L_z$ is an intrinsic value that only depends on the number of spins in the material. For the $N = 50$ cobalt atoms in the main text, the frequency of the intrinsic rotation will be around 0.23 MHz, which is one magnitude lower than the frequency of the libration (Fig. S3b).

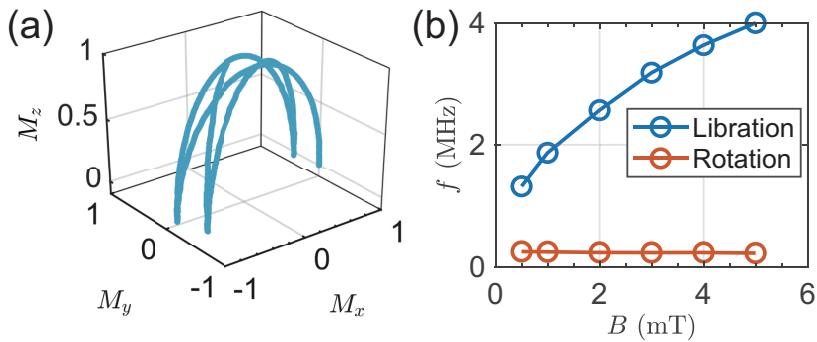

Figure S3: (a) Slowly rotation of the libration plane over ten periods. (b) Frequency of libration and the slow rotation with 50 atoms. With higher magnetic field, the libration frequency increases while the rotation frequency remains approximately the same.

## S6. METHOD FOR CALCULATING THE POWER SPECTRAL DENSITY

In this section, we outline the method for calculating the power spectral density (PSD) from a time series $\phi(t)$ obtained via dynamical simulations. According to the Wiener-Khinchin theorem [7,8], the PSD is given by the Fourier transform of the autocorrelation function of $\phi(t)$:

$$S(\omega) = \int_{-\infty}^{\infty} \langle \phi(t)\phi(t+\tau) \rangle e^{-i\omega\tau} d\tau, \tag{S9}$$

where $\langle \phi(t)\phi(t+\tau) \rangle$ is the time autocorrelation function, defined as:

$$\langle \phi(t)\phi(t+\tau) \rangle = \lim_{T \to \infty} \frac{1}{T} \int_{-T/2}^{T/2} \phi^*(t)\phi(t+\tau) dt. \tag{S10}$$

We evaluate $S(\omega)$ numerically using the fast Fourier transform (FFT). The relation between $S(\omega)$ and the squared magnitude of the Fourier transform $F(\omega)$ follows from:

$$|F(\omega)|^2 = \lim_{T\to\infty} \left| \int_{-T/2}^{T/2} \phi(t) e^{-i\omega t} \mathrm{d}t \right|^2$$

$$= \lim_{T\to\infty} \left( \int_{-T/2}^{T/2} \phi^*(t) e^{i\omega t} \mathrm{d}t \right) \left( \int_{-T/2}^{T/2} \phi(t') e^{-i\omega t'} \mathrm{d}t' \right) \quad \text{(S11)}$$

$$= \lim_{T\to\infty} T \int_{-T/2}^{T/2} \left[ \frac{1}{T} \int_{-T/2}^{T/2} \phi^*(t) \phi(t+\tau) \mathrm{d}t \right] e^{-i\omega\tau} \mathrm{d}\tau$$

$$= \lim_{T\to\infty} T S(\omega).$$

Since time series are finite in practical computations, the factor $\lim_{T\to\infty} T$ is effectively replaced by the inverse of the frequency resolution (or bandwidth) to ensure a well-defined estimate of the PSD.

For white noise $w(t)$, the power spectral density is flat, meaning $S_w(\omega) = \text{const}$. For Brownian noise, namely the noise related to Brownian motion, the corresponding time series $b(t)$ is the integral of white noise:

$$\frac{\mathrm{d}b(t)}{\mathrm{d}t} = w(t) \quad \text{(S12)}$$

By taking the Fourier transform of the Brownian noise $b(t)$ and using the integration property of the Fourier transform, we obtain:

$$F_b(\omega) = \frac{F_w(\omega)}{i\omega} \quad \text{(S13)}$$

where $F_b(\omega)$ an $F_w(\omega)$ are the FFT of Brownian noise and white noise, respectively. Therefore, the power spectral density of the Brownian noise linearly depends on $1/\omega^2$:

$$S_b(\omega) \propto |F_b(\omega)|^2 = \frac{\text{const}}{\omega^2} \quad \text{(S14)}$$

### S7. SENSITIVITY OF MEASURING AN OSCILLATING FIELD

In the main text, we proposed using nutational motion to determine the frequency of a slowly oscillating magnetic field. Here, we analyze the noise in nutation caused by intrinsic spin-lattice relaxation and evaluate the sensing sensitivity. The uncertainty in nutation over an ensemble of spins in the ferromagnetic needle is calculated as: $\Delta M_z = \sqrt{\sum_i (s_{i,z} - s_z)^2 / N}$, as Fig. S4(a) shows. The noise in nutational motion exhibits non-diffusive behavior.

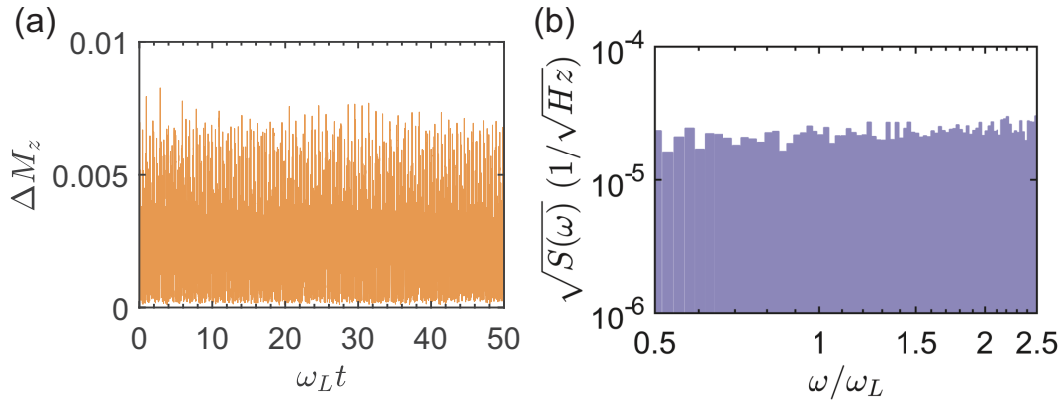

Figure S4: (a) Uncertainty of nutational motion and (b) the corresponding power spectral density

Its power spectral density, shown in Fig. S4(b), resembles that of white noise. This is similar to the precession case discussed in the main text.

Having understood the typical noise power spectrum of nutation, we demonstrate here why the uncertainty of frequency estimation decays according to $\Delta\omega \propto 1/t^{3/2}$ based on the Cramér-Rao lower bound (CRLB). Firstly, the nutational signal can be expressed as:

$$M_z(t_n) = A\cos(\omega t_n + \phi)\cos(\omega_0 t_n + \phi_0) + w[n], \tag{S15}$$

where $w[n]$ denotes Gaussian white noise with zero mean and uncertainty $\sigma$. Here, $\omega$ is the frequency of the oscillating magnetic field to be measured, $\omega_0$ is the typical oscillation frequency of nutation under a static magnetic field, $\phi$ and $\phi_0$ are arbitrary initial phase factors, and $A$ is the amplitude proportional to the moment of inertia of the needle. We consider discrete time step, where $t_n = ndt$, and $n = 0, 1, ..., N-1$. For each $t_n$, $M_z(t_n)$ is a random variable that satisfies the following likelihood function:

$$p = \frac{1}{(2\pi\sigma^2)^{N/2}}\exp\left\{-\frac{1}{2\sigma^2}\sum_{n=0}^{N-1}[M_z(t_n) - A\cos(\omega t_n + \phi)\cos(\omega_0 t_n + \phi_0)]^2\right\} \tag{S16}$$

Different from typical projective measurements that collapse quantum states, the measurement of a macroscopic ferromagnetic needle in the classical regime is weak and of non-demolition type, so that the interaction with the measurement apparatus will not destroy the state of the system. Due to this reason, we obtain a time series with a size of $N$ over the total measurement time $t_N$.

Next, we calculate the Fisher information $I(\omega)$ for the time series $M_z(t_n)$:

$$\begin{aligned} I(\omega) &= E[\frac{\partial^2 \ln p}{\partial \omega^2}] \\ &= \frac{1}{\sigma^2}\sum_{n=0}^{N-1}[\frac{\partial A\cos(\omega t_n + \phi)\cos(\omega_0 t_n + \phi_0)}{\partial \omega}]^2 \\ &\approx \frac{1}{4\sigma^2}\sum_{n=0}^{N-1} A^2 n^2 (\mathrm{d}t)^2 \\ &\approx \frac{A^2 f_{\mathrm{BW}} t_N^3}{12\sigma^2} \end{aligned} \tag{S17}$$

where we have use the summation $\sum_{n=0}^{N-1} n^2 = (N-1)N(2N-1)/6$, and $f_{\mathrm{BW}} = 1/\mathrm{d}t$ denotes frequency bandwith. According to the Cramér-Rao lower bound [9], the variance of $\omega$ must satisfy

$$\mathrm{var}(\omega) \geq \frac{1}{I(\omega)} \approx \frac{12\sigma^2}{A^2 f_{\mathrm{BW}} t_N^3}. \tag{S18}$$

Consequently, the standard deviation in estimating the frequency $\omega$ follows

$$\Delta\omega \geq \sqrt{\frac{6}{\mathrm{SNR} f_{\mathrm{BW}}^2 t_N^3}} \tag{S19}$$

where $\mathrm{SNR} = A^2/2\sigma^2$ is the signal-to-noise ratio. Thus, the sensitivity for measuring the frequency of an oscillating magnetic field using a ferromagnetic needle scales as $t^{-3/2}$ with the total time of measurement, surpassing the standard quantum limit.


\fi

\end{document}